\documentclass[pra,amsmath,amssymb,showpacs, superscriptaddress,notitlepage,nofootinbib,twocolumn]{revtex4-2}

\usepackage{amsmath,amssymb,mathtools,physics,amsfonts}
\usepackage[utf8]{inputenc}
\usepackage{amsthm}

\usepackage[colorlinks]{hyperref}
\hypersetup{
	colorlinks   = true, 
	urlcolor     = blue, 
	linkcolor    = blue, 
	citecolor   = blue 
}
\usepackage[capitalize]{cleveref}

\usepackage [english]{babel}
\usepackage [autostyle, english = american]{csquotes} 
\MakeOuterQuote{"}
\MakeInnerQuote{´}

\usepackage{xcolor}

\newif\ifmargin

\marginfalse

\ifmargin
\newdimen\extramargin
\fi
 
\newtheorem*{remark}{Theorem} 
\newtheorem*{prop}{Proposition} 

\newcommand{\affvqcc}{Vigo Quantum Communication Center, University of Vigo, Vigo E-{36310}, Spain}
\newcommand{\affuvigo}{Escuela de Ingeniería de Telecomunicación, Department of Signal Theory and Communications, University of Vigo, Vigo E-36310, Spain}
\newcommand{\affatlantic}{atlanTTic Research Center, University of Vigo, Vigo E-36310, Spain}
\newcommand{\afftoyama}{Faculty of Engineering, University of Toyama, Gofuku 3190, Toyama 930-8555, Japan}

\begin{document}
	
	\setlength{\parskip}{2pt}
	\setlength{\parindent}{0pt}
	\title{Quantum key distribution with unbounded pulse correlations}
	\author{Margarida Pereira}	\email{mpereira@vqcc.uvigo.es} \affiliation{\afftoyama}  \affiliation{\affvqcc} \affiliation{\affuvigo} \affiliation{\affatlantic} 
 \author{Guillermo Currás-Lorenzo} 
    \affiliation{\afftoyama}  \affiliation{\affvqcc} \affiliation{\affuvigo} \affiliation{\affatlantic} 
    \author{Akihiro Mizutani}
    \affiliation{\afftoyama}
    \author{Davide Rusca} 
      \affiliation{\affvqcc} \affiliation{\affuvigo} \affiliation{\affatlantic} 
    \author{Marcos Curty}
    \affiliation{\affvqcc} \affiliation{\affuvigo} \affiliation{\affatlantic} 
	\author{Kiyoshi Tamaki} 
    \affiliation{\afftoyama} 
\quad
\begin{abstract}
A prevalent issue in practical applications of quantum key distribution (QKD) is the emergence of correlations among the emitted signals. Although recent works have proved the security of QKD in the presence of this imperfection, they rest on the premise that pulse correlations are of finite length. However, this assumption is not necessarily met in practice, since the length of these correlations could be potentially unbounded. Indeed, the first emitted pulse could be correlated with the last one, even if very faintly. Still, intuitively, there should exist a pulse separation threshold after which these correlations become so small as to be essentially negligible, rendering them inconsequential from a security standpoint. Building on this insight, we introduce a general formalism designed to extend existing security proofs to the practically relevant scenario in which pulse correlations have an unbounded length. This approach significantly enhances the applicability of these proofs and the robustness of QKD's implementation security.

\end{abstract}
	\maketitle
\onecolumngrid

\section{Introduction}
Quantum key distribution (QKD) promises secure communications between two distant parties based on the laws of physics \cite{xuSecureQuantum2020,pirandolaAdvancesQuantum2020}. However, conventional security proofs of QKD often rely on idealised assumptions, neglecting inevitable device imperfections. This gap between theoretical models and real-world implementations could be exploited by an eavesdropper, compromising the security claim of QKD \cite{loSecureQuantum2014}. Addressing this challenge has become a focal point in the field \cite{zapateroImplementationSecurity2023}, with experimentalists striving to accurately characterise the magnitude of different device imperfections and refine hardware design to better match the theoretical models, and theorists developing new protocols and security proofs that accommodate various device imperfections. 

One of the most important imperfections in practice, especially among high-speed QKD systems \cite{grunenfelderPerformanceSecurity2020}, are pulse correlations. 
These occur when the setting choices made in a given round are not only encoded into the signal emitted in that round, but also inadvertently into the signals emitted in subsequent rounds. This phenomenon, purely classical in nature, can arise, for instance, from memory effects in the modulation devices. It constitutes a security risk because it could allow an eavesdropper to learn key information by investigating the leaked information in subsequent pulses, while causing no disturbance on the current one. 

Accommodating this imperfection in security proofs of QKD was believed to be difficult, as many of them require that the emitted states are independent and identically distributed \cite{xuSecureQuantum2020}. Recently, however, QKD has been proven to be secure in the presence of bit and basis correlations \cite{pereiraQuantumKey2020,mizutaniSecurityRoundrobin2021,pereiraModifiedBB842023,curras-lorenzoSecurityFramework2023}, intensity correlations \cite{zapateroSecurityQuantum2021,sixtoSecurityDecoystate2022,yoshinoQuantumKey2018} and phase-randomisation correlations \cite{curras-lorenzoSecurityQuantum2023}. 
Using these analyses, one is able to effectively bound the amount of information leaked to a potential eavesdropper and apply sufficient privacy amplification to obtain a secure key.


These proofs, however, rely on the assumption that the correlations have a finite and known maximum length $l_c$, beyond which the pulses are completely uncorrelated. In other words, one needs to guarantee that the setting choice made in the $k^{\rm th}$ round has absolutely no influence on the signal emitted in the $(k+l)^{\rm th}$ round for $l > l_c$. While it is reasonable to expect that the magnitude of the correlations decreases rapidly as the pulse separation $l$ increases, the assumption that this magnitude will drop to exactly zero for any finite value of $l$ does not seem to be justified. Indeed, these correlations could even span the entire communication sequence, i.e.~the setting choices made in the first round of the protocol could in principle influence the signals emitted in the very last round.

That being said, intuitively, there should exist a pulse separation threshold after which this influence is so small as to be almost negligible, in the sense that an eavesdropper could gain almost no information from it. This suggests that the key generated in this scenario should be almost as secure as the key that would have been generated in a scenario in which the magnitude of the correlations drops to exactly zero after the threshold. In this work, we confirm this intuition by proving that, even if the correlations technically have an unbounded length, one can apply the existing security analyses as if their length was bounded by the threshold, and then rigorously account for the neglected long-range correlations by slightly adjusting the security parameter of the final key. By doing so, we remove a significant limitation in existing security analyses, enhancing their applicability to real-world scenarios.


We remark that the formalism we introduce is rather general, and it may be used in other situations for which the existing security proofs consider a scenario that differs only slightly from the actual one. For this reason, the outline of this paper is as follows. First, in \cref{sec:protocol}, we describe a general QKD protocol. Then, in \cref{sec:main_theorem} we present our formalism for a general scenario. After that, in \cref{sec:application}, we apply it to the case of unbounded bit and basis pulse correlations and explain how experimentalists can use this result in practice. Finally, in \cref{sec:conclusion}, we summarise our findings.

\section{Description of a general QKD protocol}
\label{sec:protocol}
For clarity and simplicity, our discussion focuses on prepare-and-measure (P\&M) protocols, although our results are equally applicable to measurement-device-independent scenarios {\cite{loMeasurementDeviceIndependentQuantum2012}}. A general P\&M protocol can be described as follows: (1) Alice makes a probabilistic selection of setting choices (such as bit and basis choices) and then sends, through a quantum channel, a sequence of quantum states on systems $S_1,\hdots, S_N =: \vb*{S}$; (2) Eve performs the most general attack allowed by quantum mechanics, which, without loss of generality, can be described as the application of a unitary operator $U_{\vb*{SE}}$ on $\vb*{S}$ and on her ancillary system $\vb*{E}$, and resends the output systems $\vb*{B}$ to Bob; (3) Bob performs measurements on the received systems; (4) Alice and Bob apply post-processing (typically involving, e.g.~basis announcements, sifting, error correction, {error verification and} privacy amplification) to obtain an $\epsilon_\textrm{sec}$-secure key pair, where 
\begin{align}
    \frac{1}{2} \big|\big|  \rho_{\vb*{A'B'E'}}^{final} - \rho_{\vb*{A'B'E'}}^{ideal} \big|\big|_1  \leq \epsilon_\textrm{sec}.
    \label{eq:eps_sec}
\end{align}
Here, $\rho_{\vb*{A'B'E'}}^{final}$ is the final joint state of Alice, Bob and Eve at the end of the protocol, where $\vb*{A'}$ and $\vb*{B'}$ are Alice's and Bob's classical systems holding their respective keys {$k_A$ and $k_B$}, and $\vb*{E'}$ is Eve's ancilliary output system after applying $U_{\vb*{SE}}$. The state $\rho_{\vb*{A'B'E'}}^{ideal}$ is their joint state in an ideal protocol in which Alice and Bob share an identical key that is completely random and uncorrelated with Eve's system. Intuitively, \cref{eq:eps_sec} means that if a protocol is $\epsilon_\textrm{sec}$-secure then the probability that Eve has any information about the key and/or that Alice's and Bob's keys are not identical is at most $\epsilon_\textrm{sec}$.

\begin{figure*}[t]
    \includegraphics[width=18cm]{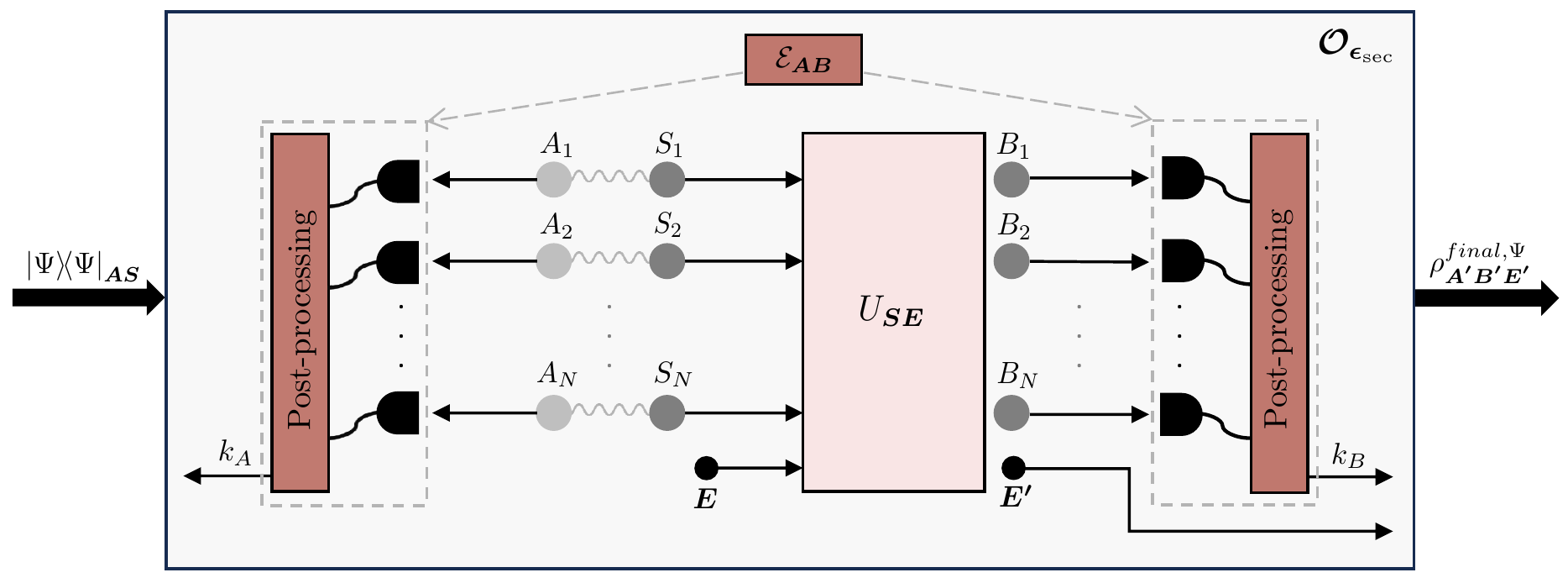} 
		\caption{Pictorial description of the quantum operation $\mathcal{O}_{\epsilon_\textrm{sec}}$, {which contains Alice's, Bob's and Eve's operations on a QKD protocol with a source replacement scheme. First, Alice prepares the entangled state $\ket{\Psi}_{\vb*{AS}}$ and sends systems $\vb*{S} = S_1, \hdots,S_N$ through the channel while keeping systems $\vb*{A} = A_1, \hdots, A_N$ in her lab. Then, Eve performs a coherent attack, which can be described by a unitary operator $U_{\vb*{SE}}$ acting on $\vb*{S}$ and Eve's ancilla system $\vb*{E}$, and resends the output systems $\vb*{B} = B_1, \hdots, B_N$ to Bob. After that, Alice and Bob perform the operation $\mathcal{E}_{\vb*{AB}}$, that is, they measure their respective systems and apply post-processing to obtain an $\epsilon_{\rm sec}$-secure key pair $k_A,k_B$. The final joint state at the end of the protocol, or in other words, after applying the quantum operation $\mathcal{O}_{\epsilon_\textrm{sec}}$, is $\rho_{\vb*{A'B'E'}}^{final,\Psi}$. }}
  \label{fig:operation}
\end{figure*}

The objective of a security analysis is proving \cref{eq:eps_sec}. To achieve this, it is often useful to assume an equivalent scenario (typically called a source replacement scheme) in which Alice generates a global entangled state $\ket{\Psi}_{\vb*{AS}}$ and then performs measurements on the ancillary systems $\vb*{A}:= A_1,\hdots,A_N$ to learn her setting choices. Also, it is helpful to consider that Alice delays her measurements until after Eve's attack. In this case, we have the following modified steps: $(1')$ Alice prepares $\ket{\Psi}_{\vb*{AS}}$ and sends systems $\vb*{S}$ through the quantum channel while keeping systems $\vb*{A}$ in her lab; $(3')$ Alice and Bob perform measurements on their local systems $\vb*{A}$ and $\vb*{B}$, respectively. We can denote Alice's and Bob's actions in steps $(3')$ and $(4)$ as a trace-preserving completely positive (TPCP) map $\mathcal{E}_{\vb*{AB}}$ such that $\mathcal{E}_{\vb*{AB}} (\hat{P} [U_{\vb*{SE}} \ket{\Psi}_{\vb*{AS}} \ket{0}_{\vb*{E}}] ) = \rho_{\vb*{A'B'E'}}^{final}$, where $\hat{P}[\cdot] = \dyad{\cdot}$. And if we define a TPCP map $\mathcal{O}_{\epsilon_\textrm{sec}}$ that also includes Eve's action in step (2), then we have that $\rho_{\vb*{A'B'E'}}^{final} = \mathcal{O}_{\epsilon_\textrm{sec}} \big(\dyad{\Psi}{\Psi}_{\vb*{AS}})$. See \cref{fig:operation} for a pictorial representation of this operation. 

\vspace{-0.1cm}
\section{Main theorem}
\label{sec:main_theorem}

Let us suppose that we have a security proof guaranteeing that, when Alice and Bob run a particular protocol in which the prepared entangled state is $\ket{\Psi}_{\vb*{AS}}$, the final key pair is {$\epsilon_{\rm sec}$-}secure. What would happen if Alice {replaces} this state by some other state that is close to it? According to our Theorem below, the resulting key pair would still be secure, albeit with a modified security parameter, {which depends on how close the two states are.}

\begin{remark}
If a QKD protocol whose prepared entangled state is $\ket{\Psi}_{\vb*{AS}}$ has been proven to be $\epsilon_\textrm{sec}$-secure, then the same protocol but whose prepared entangled state is instead $\ket{\Phi}_{\vb*{AS}}$ is $(\epsilon_{sec} + 2d)$-secure, where $d$ denotes the trace distance between $\ket{\Psi}_{\vb*{AS}}$ and $\ket{\Phi}_{\vb*{AS}}$. 
\end{remark}

\textit{Proof.} The goal is to upper bound $\frac{1}{2} \big|\big| \rho_{\vb*{A'B'E'}}^{final,\Phi} - \rho_{\vb*{A'B'E'}}^{ideal,\Phi}\big|\big|_1$, where the superscript $\Phi$ indicates the prepared entangled state,
\begin{align}
\rho_{\vb*{A'B'E'}}^{final,\Phi} &= \sum_{K\geq0} p_{\Phi}(K) \sum_{k_A,k_B = 0}^{2^K -1} p_{\Phi}(k_A,k_B|K) \dyad{k_A,k_B}{k_A,k_B}_{\vb*{A'B'}} \otimes \rho_{{\vb*{E'}}|K}^{final,\Phi} (k_A,k_B) {=:} \sum_{K\geq0} p_{\Phi}(K) \rho_{{\vb*{A'B'E'}}|K}^{final,\Phi},
\label{eq:rho_final_phi} 
\end{align}
and
\begin{align}
\rho_{\vb*{A'B'E'}}^{ideal,\Phi} &= \sum_{K\geq0} p_{\Phi}(K) \frac{1}{2^K} \sum_{k = 0}^{2^K -1} \dyad{k,k}{k,k}_{\vb*{A'B'}} \otimes \Tr_{\vb*{A'B'}} \big[\rho_{{\vb*{A'B'E'}}|K}^{final,\Phi}\big].
\label{eq:rho_ideal_phi}
\end{align}
Here, $p_{\Phi}(K)$ is the probability distribution of obtaining a final key of length $K$ and $p_{\Phi}(k_A,k_B|K)$ is the probability that Alice and Bob get the keys $k_A$ and $k_B$ given $K$. Note that in \cref{eq:rho_final_phi,eq:rho_ideal_phi} we are implicitly assuming a variable key length $K$ with $K=0$ corresponding to the case in which the protocol aborts. 

To achieve our goal, we first introduce the analogous states $\rho_{\vb*{A'B'E'}}^{final,\Psi}$ and $\rho_{\vb*{A'B'E'}}^{ideal,\Psi}$, that are defined by simply replacing $\Phi$ with $\Psi$ in \cref{eq:rho_final_phi,eq:rho_ideal_phi}, respectively. {Note that $\rho_{\vb*{A'B'E'}}^{ideal,\Phi}$ and $\rho_{\vb*{A'B'E'}}^{ideal,\Psi}$ are not equal because the reduced state on Eve's system $\vb*{E'}$ depends on whether Alice prepares $\ket{\Phi}_{\vb*{AS}}$ or $\ket{\Psi}_{\vb*{AS}}$.} Then, {by} using the triangle inequality consecutively we have that 
\begin{align}
\frac{1}{2}\big|\big| \rho_{\vb*{A'B'E'}}^{final,\Phi} - \rho_{\vb*{A'B'E'}}^{ideal,\Phi}\big|\big|_1  &\leq \frac{1}{2}\big|\big|  \rho_{\vb*{A'B'E'}}^{final,\Phi} - \rho_{\vb*{A'B'E'}}^{final,\Psi}\big|\big|_1  + \frac{1}{2}\big|\big|  \rho_{\vb*{A'B'E'}}^{final,\Psi} - \rho_{\vb*{A'B'E'}}^{ideal,\Phi}\big|\big|_1  \nonumber  \\
&\leq \frac{1}{2}\big|\big|  \rho_{\vb*{A'B'E'}}^{final,\Phi} - \rho_{\vb*{A'B'E'}}^{final,\Psi}\big|\big|_1  + \frac{1}{2}\big|\big|  \rho_{\vb*{A'B'E'}}^{final,\Psi} - \rho_{\vb*{A'B'E'}}^{ideal,\Psi}\big|\big|_1 +  \frac{1}{2}\big|\big|  \rho_{\vb*{A'B'E'}}^{ideal,\Psi} - \rho_{\vb*{A'B'E'}}^{ideal,\Phi}\big|\big|_1. 
\label{eq:trace_phi}
\end{align} 
Next, we bound each term in {the last inequality of} \cref{eq:trace_phi} separately: \\ 

 \textbf{1\textsuperscript{st} term:} As discussed in \cref{sec:protocol}, when Alice prepares the entangled state $\ket{\Psi}_{\vb*{AS}}$, the joint state of Alice, Bob and Eve at the end of the protocol can be expressed as  
    \begin{align}
    \rho_{\vb*{A'B'E'}}^{final,\Psi} = \mathcal{O}_{\epsilon_\textrm{sec}} \big(\dyad{\Psi}{\Psi}_{\vb*{AS}}).
    \label{eq:psi_operation}
    \end{align} 
{Note that this protocol is $\epsilon_{\rm sec}$-secure for \textit{any} fixed unitary operator $U_{\vb*{SE}}$, since the existing security proof did not impose any restrictions on Eve's operation, and therefore, $U_{\vb*{SE}}$ can be the operator that would have been the most advantageous to Eve if Alice had prepared the state $\ket{\Phi}_{\vb*{AS}}$ instead. If we now substitute the prepared entangled state $\ket{\Psi}_{\vb*{AS}}$ by $\ket{\Phi}_{\vb*{AS}}$, their final joint state is instead }
%
    \begin{align}
    \rho_{\vb*{A'B'E'}}^{final,\Phi} = \mathcal{O}_{\epsilon_\textrm{sec}} \big(\dyad{\Phi}{\Phi}_{\vb*{AS}}).
    \label{eq:phi_operation}
    \end{align}
    Then, by substituting \cref{eq:psi_operation,eq:phi_operation} in the first term of \cref{eq:trace_phi}, we have that
    \begin{align}
    &\frac{1}{2}\big|\big|  \rho_{\vb*{A'B'E'}}^{final,\Phi} - \rho_{\vb*{A'B'E'}}^{final,\Psi}\big|\big|_1  = T \big(\mathcal{O}_{\epsilon_\textrm{sec}} \big(\dyad{\Phi}{\Phi}_{\vb*{AS}}\big), \mathcal{O}_{\epsilon_\textrm{sec}} \big(\dyad{\Psi}{\Psi}_{\vb*{AS}} \big)\big) \leq T \big(\dyad{\Phi}{\Phi}_{\vb*{AS}},\dyad{\Psi}{\Psi}_{\vb*{AS}}\big) =: d,
    \label{eq:bound1}
    \end{align} 
where we have used the fact that the trace distance {$T(\dyad{\cdot} ,\dyad{\cdot})$} is non-increasing by quantum operations. \\ 

\textbf{2\textsuperscript{nd} term:} Since the QKD protocol is assumed to be $\epsilon_\textrm{sec}$-secure {when Alice prepares the entangled state $\ket{\Psi}_{\vb*{AS}}$}, by definition, the second term in \cref{eq:trace_phi} is bounded by $\epsilon_\textrm{sec}$ (see \cref{eq:eps_sec}). \\

\textbf{3\textsuperscript{rd} term:} The ideal states $\rho_{\vb*{A'B'E'}}^{ideal,\Phi}$ and $\rho_{\vb*{A'B'E'}}^{ideal,\Psi}$ can be directly obtained from their respective actual states $\rho_{\vb*{A'B'E'}}^{final,\Phi}$ and $\rho_{\vb*{A'B'E'}}^{final,\Psi}$ by simply replacing the actual keys $k_A$ and $k_B$ with the ideal key pair. By defining this TPCP map as $\Gamma$ (see \cref{sec:gamma} for more details), we have that the third term in \cref{eq:trace_phi} becomes 
    \begin{align}
    &\frac{1}{2}\big|\big|  \rho_{\vb*{A'B'E'}}^{ideal,\Psi} - \rho_{\vb*{A'B'E'}}^{ideal,\Phi}\big|\big|_1  = \frac{1}{2}\big|\big|  \Gamma (\rho_{\vb*{A'B'E'}}^{final,\Psi}) - \Gamma (\rho_{\vb*{A'B'E'}}^{final,\Phi}) \big|\big|_1  \leq \frac{1}{2}\big|\big|  \rho_{\vb*{A'B'E'}}^{final,\Phi} - \rho_{\vb*{A'B'E'}}^{final,\Psi}\big|\big|_1  \leq d,
    \label{eq:bound3}
    \end{align}
    where in the last inequality we have used \cref{eq:bound1}. \\

Finally, by substituting \cref{eq:bound1,eq:bound3} into \cref{eq:trace_phi} and using the fact that the protocol in which Alice prepares $\ket{\Psi}_{\vb*{AS}}$ is $\epsilon_\textrm{sec}$-secure by definition, we obtain the following bound
\begin{align}
\frac{1}{2}\big|\big|  \rho_{\vb*{A'B'E'}}^{final,\Phi} - \rho_{\vb*{A'B'E'}}^{ideal,\Phi}\big|\big|_1  \leq \epsilon_\textrm{sec} + 2d,
\end{align}
as required.

\section{Application of the theorem to unbounded pulse correlations}
\label{sec:application}

{Let} us consider a practical scenario in which Alice employs an imperfect source that introduces {bit and basis} correlations between the emitted pulses. In this case, the state of the $k^{\textrm{th}}$ pulse depends not only on Alice's $k^{\textrm{th}}$ setting choice $j_k$, but also on her previous setting choices $j_{k-1}, j_{k-2},\hdots, j_1$. {We can quantify the strength of the correlation between pulses separated by $l$ rounds, denoted by $\epsilon_l$, by considering the maximum variation that the state on the $k^{\rm th}$ round can undergo when the $(k-l)^{\rm th}$ setting choice is altered, that is,}
\begin{align}
&\big|\braket*{\psi_{j_k|j_{k-1},...,j_{k-l+1},\tilde{j}_{k-l},j_{k-l-1},...,j_1}}{\psi_{j_k|j_{k-1},...,j_{k-l+1},{j}_{k-l},j_{k-l-1},...,j_1}}\big|^2 \geq 1 - \epsilon_{l}.
\label{eq:assumption1}
\end{align}
{The} existing security proofs {addressing this imperfection}  {\cite{pereiraQuantumKey2020,mizutaniSecurityRoundrobin2021,pereiraModifiedBB842023,curras-lorenzoSecurityFramework2023}} require the assumption {that a bound on $\epsilon_l$ is known and} that the correlations have a {finite} length, i.e.~that there is a certain length $l_c$ such that $\epsilon_l = 0$ for all $l>l_c$. {The latter} condition is needed because these proofs divide the protocol rounds in $l_c+1$ groups and prove the security of each group separately, which can only be done if $l_c$ is bounded. {Unfortunately, however, while it seems natural that the strength of the correlations should decrease rapidly as the pulse separation $l$ increases, it is unreasonable to assume that it will decrease to exactly zero at any point.} 

{That being said,} there must exist a certain pulse separation $l$ after which the strength of the correlations is so small that it is essentially negligible. {Let} us denote this value of $l$ as the effective maximum correlation length $l_e$. Using {the} Theorem {in the previous section}, we can make this intuition explicit. First of all, we define the following source replacement scheme for {the} protocol: 
\begin{align}   
  &\ket{\Psi_\infty}_{\vb*{AS}} = \sum_{j_1} {\sqrt{p_{j_1}}} e^{i\theta_{j_1}} \ket*{j_1}_{A_1} \ket{\psi_{j_1}}_{S_1} \sum_{j_2} {\sqrt{p_{j_2}}} e^{i\theta_{j_1,j_2}} \ket{j_2}_{A_2} \ket*{\psi_{j_2|j_1}}_{S_2} \hdots  \sum_{j_N} {\sqrt{p_{j_N}}} e^{i\theta_{j_1,...,j_N}}\ket{j_N}_{A_N} \ket*{\psi_{j_N|j_{N-1},\hdots,j_1}}_{S_N},
    \label{eq:full_entangled_state}
 \end{align}
where {$\{\ket{j_k}_{A_k}\}_{j_k}$ is an orthonormal basis for the system $A_k$ and} the terms $e^{i\theta_{j_1,\hdots,j_k}}$ are complex phases that have no effect on Alice's measurements on systems ${\vb*{A}}$. {The} motivation to include {these phases} will be understood soon. Also, we {introduce} the following state 
\begin{align}
    &\ket{\Psi_{ l_e}}_{\vb*{AS}} = \sum_{j_1} {\sqrt{p_{j_1}}} \ket*{j_1}_{A_1} \ket{\psi_{j_1}}_{S_1} \sum_{j_2} {\sqrt{p_{j_2}}} \ket{j_2}_{A_2} \ket*{\psi_{j_2|j_1}}_{S_2} \hdots  \sum_{j_N} {\sqrt{p_{j_N}}} \ket{j_N}_{A_N} \ket*{\psi_{j_N|j_{N-1},\hdots,j_{N-l_e}}}_{S_N},
    \label{eq:full_entangled_state_lc}
 \end{align}
where we have defined 
{\begin{equation}
\ket*{\psi_{j_k|j_{k-1},\hdots,j_{k-l_e}}}_{S_k} := \ket*{\psi_{j_k|j_{k-1},\hdots,j_{k-l_e},j,j,j,...,j}}_{S_k},
\label{eq:definition}
\end{equation}}with $j,j,j,\hdots,j$ being any fixed sequence of setting choices for all rounds before the round $k-l_e$. \cref{eq:full_entangled_state_lc} represents a source replacement scheme for a fictitious scenario in which the correlations of Alice's source have a maximum bounded length of $l_e$. By applying the analyses in {\cite{pereiraQuantumKey2020,mizutaniSecurityRoundrobin2021,pereiraModifiedBB842023,curras-lorenzoSecurityFramework2023}}, one can obtain a security proof for this fictitious scenario that results in an $\epsilon_{\textrm{sec}}$-secure key. Then, provided that one can obtain the bound 
\begin{align}
T \big(\dyad{\Psi_\infty}{\Psi_\infty}_{\vb*{AS}}, \dyad{\Psi_{l_e}}{\Psi_{l_e}}_{\vb*{AS}}\big) \leq  d,
\end{align}
our Theorem ensures that, if we apply this security proof to {the} actual protocol, the final key is guaranteed to be $(\epsilon_{\textrm{sec}} + 2d)$-secure. In what follows, we first show how to bound this trace distance and then explain how to use this result in practice.

\subsection{Bounding the trace distance}
\begin{prop} The trace distance between $\ket{\Psi_{ \infty}}_{\vb*{AS}}$ and $\ket{\Psi_{ l_e}}_{\vb*{AS}}$ is bounded by
\begin{align}
T \big(\dyad{\Psi_\infty}{\Psi_\infty}_{\vb*{AS}}, \dyad{\Psi_{l_e}}{\Psi_{l_e}}_{\vb*{AS}}\big) \leq \sqrt{N \delta_{l_e}} =: d,
\end{align}
where $N$ is the number of emitted signals and $\sqrt{\delta_{l_e}} = {~\sum_{l = l_e + 1}^N \sqrt{\epsilon_l}}$.
\end{prop}

\textit{Proof.} For pure states, the trace distance can be expressed exactly in terms of their inner product as
\begin{align}
T \big(\dyad{\Psi_\infty}{\Psi_\infty}_{\vb*{AS}}, \dyad{\Psi_{l_e}}{\Psi_{l_e}}_{\vb*{AS}}\big) = \sqrt{1-|\braket{\Psi_{l_e}}{\Psi_\infty}_{\vb*{AS}}|^2}.
\label{eq:trace_dist_def}
\end{align}
Therefore, a bound on the trace distance between $\ket{\Psi_{ \infty}}_{\vb*{AS}}$ and $\ket{\Psi_{ l_e}}_{\vb*{AS}}$ can be derived by bounding $|\braket{\Psi_{l_e}}{\Psi_\infty}_{\vb*{AS}}|$. Using \cref{eq:full_entangled_state_lc,eq:full_entangled_state}, we have that 

\begin{align}
|\braket{\Psi_{l_e}}{\Psi_\infty}_{\vb*{AS}}| &= \bigg| \sum_{j_1} {p_{j_1}} e^{i\theta_{j_1}} \braket*{{\psi_{j_1}}}{{\psi_{j_1}}}_{S_1}  \hdots \sum_{j_N} {p_{j_N}} e^{i\theta_{j_1,...,j_N}} \braket*{\psi_{j_N|j_{N-1},\hdots,j_{N-l_e}}}{\psi_{j_N|j_{N-1},\hdots,j_1}}_{S_N} \bigg| \nonumber \\
&= \bigg|\sum_{j_1} {p_{j_1}} \big|\braket*{{\psi_{j_1}}}{{\psi_{j_1}}}_{S_1}\big| \hdots \sum_{j_N} {p_{j_N}}  \big|\braket*{\psi_{j_N|j_{N-1},\hdots,j_{N-l_e}}}{\psi_{j_N|j_{N-1},\hdots,j_1}}_{S_N}\big|\bigg| \nonumber \\
&= \sum_{j_1,...,j_N} {p_{j_1}} \hdots {p_{j_N}} \big|\braket*{{\psi_{j_1}}}{{\psi_{j_1}}}_{S_1}\big| \hdots   \big|\braket*{\psi_{j_N|j_{N-1},\hdots,j_{N-l_e}}}{\psi_{j_N|j_{N-1},\hdots,j_1}}_{S_N}\big| \nonumber \\
& = \sum_{j_1,...,j_N} {p_{j_1}} \hdots {p_{j_N}} \prod_{k=l_e+2}^N \big|\braket*{\psi_{j_k|j_{k-1},\hdots,j_{k-l_e}}}{\psi_{j_k|j_{k-1},\hdots,j_1}}_{S_k}\big|,
\label{eq:ip}
\end{align}

where, {without loss of generality}, we have exploited the freedom to introduce and choose the phases in \cref{eq:full_entangled_state} such that all inner products are real and positive, {i.e. $\theta_{j_1, \hdots, j_k} = -\arg(\braket*{\psi_{j_k|j_{k-1},\hdots,j_{k-l_e}}}{\psi_{j_k|j_{k-1},\hdots,j_1}}_{S_k})$}. Also, in the first equality of \cref{eq:ip} we have used $\braket{j_k}{j_k'}_{A_k} = \delta_{j_k,j_k'}$ and in the last equality we have used the fact that the first $l_e +1$ inner products equal one. 

Now, to bound the terms $|\braket*{\psi_{j_k|j_{k-1},\hdots,j_{k-l_e}}}{\psi_{j_k|j_{k-1},\hdots,j_1}}_{S_k}|$ in \cref{eq:ip} we use the definition {in \cref{eq:definition}} and exploit the relationship between trace distance and fidelity such that
\begin{align}
&\big|\braket{\psi_{j_k|j_{k-1},\hdots,j_{k-l_e}}}{\psi_{j_k|j_{k-1},\hdots,j_1}}_{S_k}\big| = \big|\braket{\psi_{j_k|j_{k-1},\hdots,j_{k-l_e},j,j,j,...,j}}{\psi_{j_k|j_{k-1},...,j_{k-l_e}, j_{k-l_e-1},j_{k-l_e-2},j_{k-l_e-3},...,j_1}}_{S_k}\big| \nonumber \\
&= \sqrt{1 - T\left(\hat{P}\big(\ket{\psi_{j_k|j_{k-1},...,j_{k-l_e},j,j,j,...,j}}_{S_k}\big),\hat{P}\big(\ket{\psi_{j_k|j_{k-1},...,j_{k-l_e}, j_{k-l_e-1},j_{k-l_e-2},j_{k-l_e-3},...,j_1}}_{S_k}\big)\right)^2}.
\label{eq:trace_fid}
\end{align} 
The trace distance term in \cref{eq:trace_fid} can be bounded as follows
\begin{align}
&T\left(\hat{P}\big(\ket{\psi_{j_k|j_{k-1},...,j_{k-l_e},j,j,j,...,j}}_{S_k}\big),\hat{P}\big(\ket{\psi_{j_k|j_{k-1},...,j_{k-l_e}, j_{k-l_e-1},j_{k-l_e-2},j_{k-l_e-3},...,j_1}}_{S_k}\big)\right) \nonumber \\
& \leq T \left(\hat{P}\big(\ket{\psi_{j_k|j_{k-1},...,j_{k-l_e},j,j,j,...,j}}_{S_k}\big),\hat{P}\big(\ket{\psi_{j_k|j_{k-1},...,j_{k-l_e}, j_{k-l_e-1},j,j,...,j}}_{S_k}\big)\right) \nonumber \\
&+ T \left(\hat{P}\big(\ket{\psi_{j_k|j_{k-1},...,j_{k-l_e},j_{k-l_e-1},j,j,...,j}}_{S_k}\big),\hat{P}\big(\ket{\psi_{j_k|j_{k-1},...,j_{k-l_e}, j_{k-l_e-1},j_{k-l_e-2},j_{k-l_e-3},...,j_1}}_{S_k}\big)\right) \nonumber \\
& \leq T \left(\hat{P}\big(\ket{\psi_{j_k|j_{k-1},...,j_{k-l_e},j,j,j,...,j}}_{S_k}\big),\hat{P}\big(\ket{\psi_{j_k|j_{k-1},...,j_{k-l_e}, j_{k-l_e-1},j,j,...,j}}_{S_k}\big)\right) \nonumber \\
&+ T \left(\hat{P}\big(\ket{\psi_{j_k|j_{k-1},...,j_{k-l_e},j_{k-l_e-1},j,j,...,j}}_{S_k}\big),\hat{P}\big(\ket{\psi_{j_k|j_{k-1},...,j_{k-l_e}, j_{k-l_e-1},j_{k-l_e-2},j,...,j}}_{S_k}\big)\right) \nonumber \\
&+ T \left(\hat{P}\big(\ket{\psi_{j_k|j_{k-1},...,j_{k-l_e},j_{k-l_e-1},j_{k-l_e-2},j,...,j}}_{S_k}\big),\hat{P}\big(\ket{\psi_{j_k|j_{k-1},...,j_{k-l_e}, j_{k-l_e-1},j_{k-l_e-2},j_{k-l_e-3},...,j_1}}_{S_k}\big)\right) \nonumber \\
&\leq T \left(\hat{P}\big(\ket{\psi_{j_k|j_{k-1},...,j_{k-l_e},j,j,j,...,j}}_{S_k}\big),\hat{P}\big(\ket{\psi_{j_k|j_{k-1},...,j_{k-l_e}, j_{k-l_e-1},j,j,...,j}}_{S_k}\big)\right) \nonumber \\
&+ T \left(\hat{P}\big(\ket{\psi_{j_k|j_{k-1},...,j_{k-l_e},j_{k-l_e-1},j,j,...,j}}_{S_k}\big),\hat{P}\big(\ket{\psi_{j_k|j_{k-1},...,j_{k-l_e}, j_{k-l_e-1},j_{k-l_e-2},j,...,j}}_{S_k}\big)\right) \nonumber \\
&+ T \left(\hat{P}\big(\ket{\psi_{j_k|j_{k-1},...,j_{k-l_e},j_{k-l_e-1},j_{k-l_e-2},j,...,j}}_{S_k}\big),\hat{P}\big(\ket{\psi_{j_k|j_{k-1},...,j_{k-l_e}, j_{k-l_e-1},j_{k-l_e-2},j_{k-l_e-3},...,j}}_{S_k}\big)\right) + \hdots  \nonumber \\
& + T \left(\hat{P}\big(\ket{\psi_{j_k|j_{k-1},...,j}}_{S_k}\big),\hat{P}\big(\ket{\psi_{j_k|j_{k-1},...,j_1}}_{S_k}\big)\right) \nonumber \\
&= \sqrt{1-\big|\braket{\psi_{j_k|j_{k-1},...,j_{k-l_e},j,j,j,...,j}}{\psi_{j_k|j_{k-1},...,j_{k-l_e}, j_{k-l_e-1},j,j,...,j}}_{S_k}\big|^2} + \hdots + \sqrt{1-\big|\braket{\psi_{j_k|j_{k-1},...,j}}{\psi_{j_k|j_{k-1},...,j_1}}_{S_k}\big|^2} \nonumber \\
& \leq \sqrt{\epsilon_{l_e +1}} + \hdots + \sqrt{\epsilon_{l_e + k-1}} = \sum_{n=1}^{k-1} \sqrt{\epsilon_{l_e+n}} \leq  \sum_{n=1}^{{N-l_e}} \sqrt{\epsilon_{l_e+n}} {~= \sum_{l = l_e + 1}^N \sqrt{\epsilon_l}} =: \sqrt{\delta_{l_e}},
\label{eq:sqrt_delta}
\end{align}
where we have used the triangle inequality consecutively. Also, in the equality of \cref{eq:sqrt_delta}, we have used the relationship between trace distance and fidelity, and in the second to last inequality of \cref{eq:sqrt_delta} we have used \cref{eq:assumption1}. Substituting 
\cref{eq:sqrt_delta} in \cref{eq:trace_fid}, we have that 
\begin{align}
\big|\braket{\psi_{j_k|j_{k-1},\hdots,j_{k-l_e}}}{\psi_{j_k|j_{k-1},\hdots,j_1}}_{S_k}\big| \geq \sqrt{1-\delta_{l_e}}.
\label{eq:ip_delta}
\end{align}
Then, substituting \cref{eq:ip_delta} in \cref{eq:ip}, we obtain
\begin{align}
|\braket{\Psi_{l_e}}{\Psi_\infty}_{\vb*{AS}}| &\geq \sum_{j_1,...,j_N} {p_{j_1}} \hdots {p_{j_N}} \prod_{k=l_e+2}^N \sqrt{1-\delta_{l_e}}  = \prod_{k=l_e+2}^N \sqrt{1-\delta_{l_e}} = (1-\delta_{l_e})^\frac{N-l_e-2}{2},
\label{eq:ip2}
\end{align}
since the probabilities sum to one. Finally, by substituting \cref{eq:ip2} in \cref{eq:trace_dist_def} and using Bernoulli's inequality, we {find that}
\begin{align}
T \big(\dyad{\Psi_\infty}{\Psi_\infty}_{\vb*{AS}}, \dyad{\Psi_{l_e}}{\Psi_{l_e}}_{\vb*{AS}}\big) &\leq \sqrt{(N-l_e-2)\delta_{l_e}} \leq \sqrt{N \delta_{l_e}} =: d,
\label{eq:final_bound}
\end{align}
as required.

\subsection{{Specific pulse correlations model}}
{To apply the Theorem in practice, one needs to determine $\delta_{l_e}$, which depends on the magnitude of the correlations $\epsilon_{l}$. As a particular example, we shall assume that this quantity decreases exponentially with the pulse separation $l$, that is, 
\begin{align}
\epsilon_{l} = \epsilon_1 e^{-C(l-1)},
\label{eq:exp_model}
\end{align}
where $\epsilon_1$ is the magnitude of nearest neighbour {pulse} correlations and $C$ is a constant that determines how fast the magnitude of the correlations drops as the separation between the pulses increases. However, our formalism could be straightforwardly adapted to other models. Using \cref{eq:exp_model},} we have that $\sqrt{\delta_{l_e}}$ can be expressed as
\begin{align}
\sqrt{\delta_{l_e}} &= \sum_{l = l_e + 1}^N \sqrt{\epsilon_l} \leq \sum_{l = l_e + 1}^\infty \sqrt{\epsilon_l} = \sum_{l = l_e + 1}^\infty \sqrt{\epsilon_1 e^{-C(l-1)}} = \frac{\sqrt{\epsilon_1 e^{-C l_e}}}{1 - \sqrt{e^{-C}}},
\label{eq:delta_lc}
\end{align}
where we have substituted \cref{eq:exp_model} in \cref{eq:sqrt_delta}. Then, by substituting \cref{eq:delta_lc} in \cref{eq:final_bound}, $d$ can be re-defined as
\begin{align}
d = \frac{\sqrt{N {\epsilon_1} e^{-Cl_e}}}{1-\sqrt{e^{-C}}}.
\end{align}
To have a good security guarantee, we want that $d$ is of the same order of magnitude as $\epsilon_\textrm{sec}$. To achieve this for a particular value of $N$, we need to appropriately choose the effective maximum correlation length $l_e$, which our security proof is based on. For this, it is useful to express $l_e$ as a function of $d$ and $N$ as 
\begin{align}
l_e = \frac{1}{C} \ln(\frac{N\epsilon_1}{d^2 (1- \sqrt{e^{-C}})^2}).
\label{eq:lc}
\end{align} 

Therefore, in practice, to prove the security of a QKD protocol {with a fixed $N$} whose pulses are all correlated one should do the following: (1) infer from {a source-characterisation} experiment the {value of the parameters $\epsilon_l$ (if they follow the expression given by \cref{eq:exp_model}, this reduces to determining the parameters $C$ and $\epsilon_1$)}; (2) decide the desired value of $d$ and calculate the effective maximum correlation length $l_e$ {(in the case of an exponential decrease, this can be done using \cref{eq:lc})}; (3) apply one of the security analyses in {\cite{pereiraQuantumKey2020,mizutaniSecurityRoundrobin2021,pereiraModifiedBB842023,curras-lorenzoSecurityFramework2023}} assuming that the true maximum correlation length $l_c$ equals $l_e$; and (4) increase the security parameter $\epsilon_\textrm{sec}$ claimed by the applied analysis by $2d$.


\subsection{Simulations for different values of $C$}
\begin{figure}[h!]
    \includegraphics[width=8.5cm]{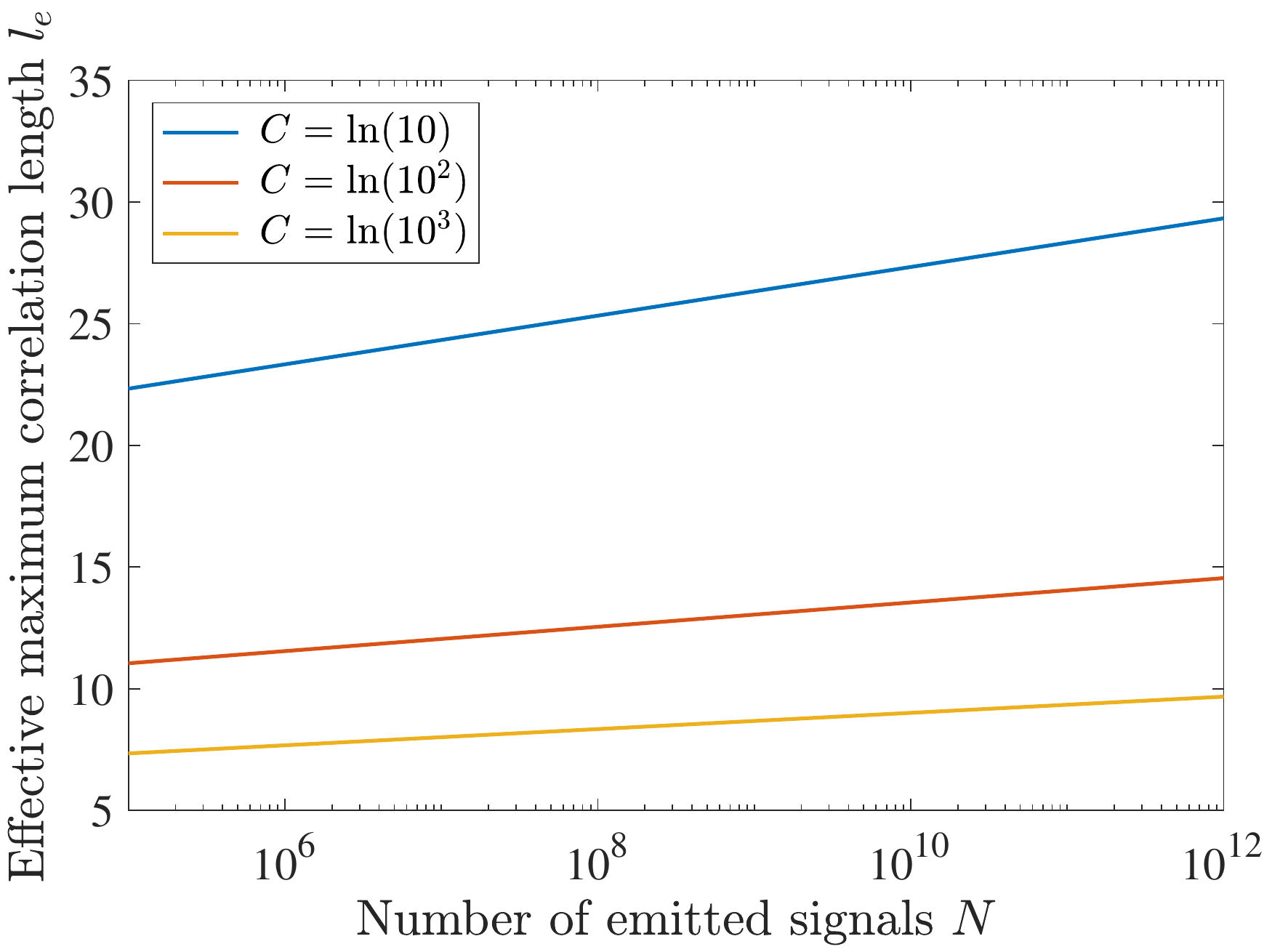} 
		\caption{Value of the effective maximum correlation length $l_e$ that one should set to achieve $d = 10^{-10}$ \cite{navarreteImprovedFinitekey2022} as a function of the number of emitted signals $N$. For the simulations, we have assumed that $\epsilon_1 = 10^{-3}$ \cite{grunenfelderPerformanceSecurity2020}.} 
  \label{fig:graph}
	\end{figure}

As a particular example, in \cref{fig:graph}, we plot the required value of $l_e$ as a function of $N$ using \cref{eq:lc}. Since to {the best of} our knowledge there are no experimental works quantifying $C$, in our simulations we consider a range of values for this parameter. Moreover, we assume that $\epsilon_1 = 10^{-3}$ \cite{grunenfelderPerformanceSecurity2020}, {and given that $10^{-10}$ is a typical value for $\epsilon_{\rm sec}$  \cite{navarreteImprovedFinitekey2022}, we assume that $d = 10^{-10}$.}

The results in \cref{fig:graph} show that as $N$ increases, $l_e$ also increases. This is expected because a larger $N$ means that potentially more pulses could be correlated with one another, and therefore one would need to set a higher $l_e$ to achieve the same level of security. {While increasing $N$ is known to reduce finite key effects, our work shows that it also leads to a higher $l_e$, thereby presenting a compromise due to the additional time required for post-processing. Moreover, in \cref{fig:graph},} one can see that the parameter $C$, which quantifies how fast the magnitude of the correlations drops with distance, has a high impact on the required $l_e$. Again, this is expected because if $C$ drops very fast then the correlations between far-away pulses will be very faint, allowing us to achieve the desired level of security with a smaller {value of} $l_e$. 

\section{Conclusion}
\label{sec:conclusion}

Quantum key distribution (QKD) implementations often suffer from correlations among the emitted signals. Recently, QKD has been shown to be secure in {their presence} {\cite{pereiraQuantumKey2020,mizutaniSecurityRoundrobin2021,pereiraModifiedBB842023,curras-lorenzoSecurityFramework2023, zapateroSecurityQuantum2021,sixtoSecurityDecoystate2022,curras-lorenzoSecurityQuantum2023,yoshinoQuantumKey2018}}. However, these analyses require the assumption that the correlations have a bounded length, which is not necessarily the case in practice. In this work, we have provided a {general} formalism to extend {existing} security proofs to the more realistic scenario in which the length of the correlations may be unbounded. 

Our approach involves the consideration of an effective maximum correlation length $l_e$, which should be chosen such that the magnitude of the residual correlations between pulses separated by more than $l_e$ rounds is so small as to be almost negligible. Here, by "almost negligible", we mean that the global entangled state prepared in the actual protocol cannot be distinguished from the global entangled state that would have been prepared in a protocol for which the magnitude of these residual correlations is exactly zero, except with a tiny failure probability $d$. More specifically, we have shown that, under this condition, one can simply 
apply the {existing} security proofs {\cite{pereiraQuantumKey2020,mizutaniSecurityRoundrobin2021,pereiraModifiedBB842023, curras-lorenzoSecurityFramework2023,zapateroSecurityQuantum2021,sixtoSecurityDecoystate2022,curras-lorenzoSecurityQuantum2023}} as if the true maximum correlation length was indeed $l_e$, and then account for the residual correlations beyond this limit by simply increasing the security parameter of the final key by $2d$. 

To show how one can apply our formalism, {we have focused on the scenario in which the emitted signals suffer from bit and basis correlations, which was considered in {\cite{pereiraQuantumKey2020,mizutaniSecurityRoundrobin2021,pereiraModifiedBB842023,curras-lorenzoSecurityFramework2023}}. For this, we have assumed that the} magnitude of the correlations decreases exponentially with their length, and used it to determine the appropriate value of $l_e$ as a function of the total number of transmitted rounds $N$, the desired failure probability $d$, and the exponential decay constants. {We remark, however, that} our formalism can also be applied to extend security proofs addressing intensity correlations \cite{zapateroSecurityQuantum2021,sixtoSecurityDecoystate2022} and phase-randomisation correlations \cite{curras-lorenzoSecurityQuantum2023} to the case in which these correlations have an unbounded length. Our result significantly increases the practical applicability of these security analyses, and advances the state of the art of QKD's implementation security.





\section{Acknowledgements} 
We thank Ainhoa Agulleiro, Fadri Grünenfelder, Ana Blázquez and {Xoel Sixto} for valuable discussions. This work was supported by Cisco Systems Inc.,
the Galician Regional Government (consolidation of Research Units: AtlantTIC), the Spanish Ministry of
Economy and Competitiveness (MINECO), the Fondo
Europeo de Desarrollo Regional (FEDER) through
the grant No.~PID2020-118178RB-C21, MICIN with
funding from the European Union NextGenerationEU
(PRTR-C17.I1) and the Galician Regional Government
with own funding through the “Planes Complementarios de I+D+I con las Comunidades Autónomas” in
Quantum Communication, the European Union’s Horizon Europe Framework Programme under the Marie
Sklodowska-Curie Grant No.~101072637 (Project QSI)
and the project “Quantum Security Networks Partnership” (QSNP, grant agreement No.~101114043). M.P.~and G.C.-L.~acknowledge support from JSPS Postdoctoral Fellowships for Research in Japan. K.T.~acknowledges support from JSPS KAKENHI Grant Number 23H01096.
\bibliography{Refs}

\appendix 
    
\section{Constructing $\Gamma$}
\label{sec:gamma}
The construction of $\Gamma$ is as follows. First note that the final state can be expressed as 
\begin{align}
\rho_{\vb*{A'B'E'}}^{final} &= \sum_{K\geq0} p(K) \sum_{k_A,k_B = 0}^{2^K -1} p(k_A,k_B|K) \dyad{k_A,k_B}{k_A,k_B}_{\vb*{A'B'}} \rho_{{\vb*{E'}}|K}^{final} (k_A,k_B) \nonumber \\
& = \Tr_K \sum_{K\geq0} p(K) \dyad{K}{K}_{K} \sum_{k_A,k_B = 0}^{2^K -1} p(k_A,k_B|K) \dyad{k_A,k_B}{k_A,k_B}_{\vb*{A'B'}} \rho_{{\vb*{E'}}|K}^{final} (k_A,k_B).
\label{eq:firstApp}
\end{align}
Then, by taking the trace over $\vb*{A'B'}$ we obtain 
\begin{align}
\Tr_K \sum_{K\geq0} p(K) \dyad{K}{K}_{K} \sum_{k_A,k_B = 0}^{2^K -1} p(k_A,k_B|K) \rho_{{\vb*{E'}}|K}^{final} (k_A,k_B), 
\end{align}
and after adding the state $\ket{0}_{\vb*{A'B'}}$ we arrive to 
\begin{align}
\Tr_K \sum_{K\geq0} p(K) \dyad{K}{K}_{K} \dyad{0}{0}_{\vb*{A'B'}} \sum_{k_A,k_B = 0}^{2^K -1} p(k_A,k_B|K) \rho_{{\vb*{E'}}|K}^{final} (k_A,k_B).
\end{align}
Finally, we swap the state of $\vb*{A'B'}$ with the ideal key state $\tau_K := 1/2^K \sum_{k=0}^{2^K -1} \dyad{k,k}{k,k}_{\vb*{A'B'}}$ by controlling system $K$, leading to
\begin{align}
&\Tr_K \sum_{K\geq0} p(K) \dyad{K}{K}_{K} \tau_K \sum_{k_A,k_B = 0}^{2^K -1} p(k_A,k_B|K) \rho_{{\vb*{E'}}|K}^{final} (k_A,k_B) \nonumber \\
&= \sum_{K \geq 0} p(K) \frac{1}{2^K} \sum_{k = 0}^{2^K -1} \dyad{k,k}{k,k}_{\vb*{A'B'}} \sum_{k_A,k_B = 0}^{2^K -1} p(k_A,k_B|K) \rho_{{\vb*{E'}}|K}^{final} (k_A,k_B) = \rho_{\vb*{A'B'E'}}^{ideal}.
\label{eq:lastApp}
\end{align}
The transformation from \cref{eq:firstApp} to \cref{eq:lastApp}, which we call $\Gamma$, is a TPCP map that takes the actual state $\rho_{\vb*{A'B'E'}}^{final}$ into its respective ideal state $\rho_{\vb*{A'B'E'}}^{ideal}$.

\end{document}